# Supercontinuum Generation in Photonic Crystal Fibers Possessing High Birefringence and Large Optical Nonlinearity

**Mohit Sharma*[1], Nitu Borgohain[2], S. Konar[3]**

[1, 2, 3]Department of Physics, Birla Institute of Technology, Ranchi 835215, India

Email address – mohitsharmac@gmail.com

## Abstract

This paper presents the design of an index guided highly birefringent photonic crystal fiber which promises to yield very large birefringence ($\sim 3.33 \times 10^{-2}$) at 1550nm and ($\sim 1.75 \times 10^{-2}$) at 1064nm as well as large effective nonlinearity ($\sim 80\ W^{-1}\ km^{-1}$). Optical supercontinuum generation in the proposed fiber using a 1064 nm pump source with peak power of 1kW has been also presented. Finite difference time domain method (FDTD) has been employed to examine the optical properties such as fiber birefringence, mode field, V-parameter, walk-off and optical nonlinearity, while the Split-step Fourier method is used to solve the nonlinear Schrödinger equation felicitating the study of supercontinuum generation. Simulation results indicate that horizontal input pulse yields superior continuum in comparison to that of the vertically polarized input. However, the broadening of the continuum is about 1450 nm in case of horizontally polarized input light whereas it is approximately 2350 nm for vertically polarized.

1. **Introduction**

Photonic crystal fiber (PCF) technology has rapidly progressed in recent years and attracted much attention of researchers [1-10]. In general, they are made from a single material like silica with microscopic air holes running along the length of the fiber. Modification of several parameters such as air hole arrangement, air hole pitch and diameter, air hole shape, refractive index of the fiber material, allows to obtain desirable optical properties like endlessly single mode operation [5-17] tailorable dispersion [5-17], large optical nonlinearity [18-19], high birefringence [21-33], large mode area [20] etc. Fiber birefringence is an important optical property, which has been exploited in coherent optical communications as well as in the fabrication of fiber optic sensors [21-30]. In conventional optical fibers, birefringence is achieved either by introducing stress in the fiber profile or by creating asymmetry in the core region. However, in conventional optical fibers, achievable birefringence is small due to small index contrast. Contrary to this, high birefringence can be easily achieved in PCFs due to superior design flexibility and large index contrast. Usually high birefringence is achieved in PCFs either by designing asymmetric core, or change of the size of a few selected air holes [21, 22] or by using squeeze crystal lattice in which number of air holes along two orthogonal axes is different [23]. Introduction of elliptical air holes is also a powerful technique to achieve high birefringence [24-26].

Theoretical as well as experimental investigations have been carried out by several authors to examine the birefringence properties of photonic crystal fibers [22-33]. Birefringence properties of band gap guiding asymmetric core PCFs have been studied by Saitoh *et al.* [27]. Birefringence $\sim 10^{-3}$ at wavelength $1.55\ \mu m$ in band gap guiding PCFs has been reported experimentally [26]. J. Wang *et al.* [29] have proposed a band gap guiding PCFs with elliptical air holes. Recently, Lyngso *et al.* [30] have reported two different types of polarization maintaining solid core PCFs that guide light by a combination of a photonic band gap and total internal reflection. They have studied group and phase birefringence experimentally and numerically, whose value is of the order of $(\sim 10^{-4})$. Cho *et al.* [31] have reported highly birefringent terahertz polarization maintaining plastic PCFs which exhibit an extremely large birefringence. Recently, Rakhi et. al. [32] have reported very large birefringence $(\sim 5.45 \times 10^{-2})$ in a band gap guided fiber, while Mohit et. al. [33] have also reported the birefringence ($\sim 2.2 \times 10^{-2}$) in index guiding PCFs at 1.55 µm. Recently, several authors have fabricated the PCFs with air holes pitch $\Lambda < 1\,\mu m$. For example, Francisco *et. al.* [34] have demonstrated the fabrication of a PCF with $d/\Lambda = 0.89$ and $\Lambda = 0.93\,\mu m$, and Magi *et. al.* [35] have also fabricated the tapered PCFs with $\Lambda \sim 0.56\,\mu m$ by using flame brushing technique [36,37]. Therefore, it appears that the proposed fiber may be fabricated without much difficulty.

Among diverse applications of PCFs, one most popular is the generation of supercontinuum (SC) [38-40]. Owing to large index contrast, PCFs allow much stronger mode confinement leading to large effective nonlinearity which is the most important factor governing the supercontinuum dynamics. Due to large effective optical nonlinearity, the threshold power requirement is low, thereby making PCFs most popular and attractive medium for supercontinum generation. SC generation in PCFs has become extremely popular because of its diverse applications such as optical coherence tomography [41,42], optical metrology [43,44], frequency comb generation [45,46], time resolved applications and spectroscopy [47,48]. The significant amount of research works have been carried out to understand the underlying mechanisms behind the broadband spercontinuum generation [38-54]. SC spectra are usually generated by pumping femto-second or pico-second pulses in the highly nonlinear PCF with anomalous group velocity dispersion (GVD) regime of the fiber, close to the zero-dispersion wavelength. In this case, spectral broadening is contributed by a hoast of nonlinear phenomena such as self-phase modulation, modulation instability, Raman scattering together with third and higher order dispersion. This process produces a spectrum with a fine and complex structure, which is very sensitive to pump pulse fluctuations. For optical communications the SC lightwave should possess low noise and high stability. In order to reduce noise and increase stability, SC can be generated in the normal GVD regime where spectral broadening is mainly dominated through self phase modulation, cross phase modulation, Raman scattering and four wave mixing [49,50]. Considerable efforts have been made to study SC in PCFs with a single zero dispersion wavelength [40]. Several authors have directed their efforts to achieve enhanced frequency shifting in the shorter wavelength [51, 52] which is possible only when appropriate group index matching is satisfied [53, 54]. Enhanced SC bandwidth with improved flatness has been demonstrated using PCFs with two zero dispersion wavelengths [55-57]. In addition, recently it has been shown that multi wavelength pumping also allows for significant enhancement of the SC bandwidth [58, 59]. Due to large optical nonlinearity, non-silica fibers have been exploited to generate supercontinuum in the wavelength range 2-5 $\mu m$ [60-64]. Solid core photonic band gap fibers have been also investigated to generate supercontinuum in the visible region [65]. Supercontinuum generation due to slow nonlinear response in a liquid filled photonic crystal fiber and modulation instability induced SC due to saturable nonlinear response have been also examined [66-68]. Towards this direction, we concentrate our efforts to investigate the SC generation in highly birefringent silica PCFs. Therefore, in this paper, we present the design of a PCF which promises to yield very large birefringence. To be compatible for SC generation, the fiber is designed to yield large optical nonlinearity which is arising due to strong mode confinement. We exploit the fiber nonlinearity to demonstrate that the fiber could be useful in supercontinuum generation. The properties of generating SC have been examined. The novelty of the present investigation lies in the fact that the studied fiber simultaneously posses large birefringence and optical nonlinearity. Moreover, it is capable of generating flat supercontinuum at low power level.

## 2.1. Fiber Modeling Method

Several modeling techniques have been applied to characterize PCFs, after its first experimental demonstration in 1996 [1]. These are plane-wave expansion method [69-71], localized-function method [72, 73], approximate effective index model [74, 75], finite element method [76], multipole method [77, 78], finite difference frequency domain (FDFD) method [79] and finite difference time domain (FDTD) method [80, 81]. A detailed comparison of these methods have been done by Gallagher [82]. In this paper, we use finite difference time domain (FDTD) technique to analyze the optical properties of photonic crystal fibers. The FDTD method is fully capable to describe arbitrary structure, user friendly, well established and tested and a wide range of wavelength response may be achieved in a single simulation. FDTD is a time-domain process, which calculates the H and E electromagnetic fields at each point of computational domain and provides real time graphical representation of electromagnetic field propagation in the fiber.

In a conventional 3-Dimension FDTD method [16, 17], the Yee's algorithm [16] is implemented to discrete Maxwell's equation for both electric and magnetic fields in a cartesian coordinate. Basically, there are three steps to solve a 3-Dimension arbitrary structure by FDTD method. First, each E component surrounded by four H components and each H component is surrounded by four E components. Second, the leapfrog time-step adopted, where all the E components in the 3-Dimension spaces are calculated as well as stored in memory for a particular time point using the H data previously stored in memory. Then all the H data are calculated as well as stored in memory using the E data just computed. And then finally, all derivatives in Maxwell's equations, including the space derivatives and time derivatives, are replaced by the central finite difference. The FDTD method allows to specify the materials at all points within the gridded computational domain. This method also allow to choose a wide variety of linear and nonlinear dielectric and magnetic materials, which can be easily modeled.

Since FDTD requires that the entire computational domain be gridded, and the grid spatial discretization must be sufficiently fine to resolve both the smallest electromagnetic wavelength and the smallest geometrical feature in the model, very large computational domains can be developed, which results in very long solution times. Models with long, thin features, (like wires) are difficult to model in FDTD because of the excessively large computational domain required. Methods such as Eigenmode Expansion can offer a more efficient alternative as they do not require a fine grid along the z-direction [82].

The effective index of the fundamental optical mode inside fiber has been calculated using $n_{eff} = \frac{\lambda}{2\pi}\beta$, where $\beta$ is the propagation constant. When a linearly polarized light is sent through the core of an anisotropic fiber, there is a difference in refractive index in two mutually perpendicular directions, hence, an entering light wave is split into two different orthogonal polarized modes each moving with different velocity depending on the refractive index in the respective direction. This difference in refractive indices is called as birefringence. The modal birefringence [2] is the difference of the real part of the effective refractive indices of two modes in x and y-directions i.e., $B = |Re\,(n_{eff}^x) - Re\,(n_{eff}^y)|$, where Re represents the real part, $n_{eff}^x$ and $n_{eff}^y$ are the effective refractive indices of optical modes polarized along x and y-directions, respectively. Fiber dispersion consists of two components, one is the waveguide dispersion, which arises due to guiding and confinement of electromagnetic waves in the fiber, while the other is due to the fiber material and is known as material dispersion. The waveguide dispersion is calculated from the effective index using the relationship $D_w(\lambda) = -\frac{\lambda}{c}\frac{d^2 n_{eff}}{d\lambda^2}$, where c is the velocity of light in vacuum[3]. Material dispersion $D_m(\lambda) = -\frac{\lambda}{c}\frac{d^2 n}{d\lambda^2}$ can be estimated using Sellmier's [32] equation $n^2(\omega) = 1 + \sum_{j=1}^{m}\frac{B_j \omega_j^2}{\omega_j^2 - \omega^2}$. For bulk fused silica $B_1 = 0.6961663$, $B_2 = 0.4079426$, $B_3 = 0.8974794$, $\lambda_1 = 0.0684043\ \mu m$, $\lambda_2 = 0.1162414\ \mu m$ and $\lambda_3 = 9.896161\ \mu m$, where $\lambda_j = \frac{2\pi c}{\omega_j}$, $\omega_j$ is the resonance frequency. An important parameter which characterizes the single modeness of PCFs is the normalized V parameter whose effective value $V_{eff}$ can be expressed [1] as $V_{eff} = 2\pi\frac{\Lambda}{\lambda}\sqrt{n_c^2 - n_{eff}^2}$, where $\Lambda$ is the hole pitch, $\lambda$ is the wavelength in vacuum, $n_c$ is the refractive index of the core. The single mode cut-off for photonic crystal fiber is $V_{eff} \leq 4.1$. The nonlinear coefficient of the PCF is another important parameter characterized b y $= \frac{2\pi n_2}{\lambda A_{eff}}$, where $n_2$ is the nonlinear Kerr coefficient of the fiber material and $A_{eff}$ is the effective mode field area. The value of the nonlinear coefficient can be maximized either by tailoring effective area or selecting a material with large $n_2$ or by manipulating both. In the present investigation, we use silica PCFs which are characterized by small value of $n_2$ and optimize fiber design parameters to achieve small effective area, thus leading to large nonlinearity.

## 2.2. Fiber Structure

One of the main objectives of the present investigation is to evolve certain fiber designs which promise to yield large birefringence ($\sim 3.33 \times 10^{-2}$). The key idea is to destroy the symmetry of the PCF structure, which increases the difference between the refractive indices of the two mutually polarized modes. In order to achieve this, we have designed elliptical core photonic crystal fiber. In Figure 1 (a) we have displayed the cross section of the fiber. It composed of the seven rings of air holes of diameter $d$ and hole pitch $\Lambda$, which are arranged in a triangular lattice formation. The hole diameter of four outer rings is same, while hole diameter of three innermost rings is different. Several holes of the first, second and third rings have been joined together to form two large elliptical air holes of ellipticity $\eta$, where ellipticity may be defined as $\eta = \frac{d_a}{d_b}$, $d_a$ and $d_b$ are the lengths of the major and minor axes, respectively. Two holes each from first and second rings have been omitted to form the elliptic core of the fiber. In the third ring two diagonally opposite circular air holes of diameter $d_l$ have been made larger to ensure optical confinement. These two air holes play crucial role in optical confinement and their role can be appreciated examining figure 1(b) carefully. In this figure, the diameter of these two air holes has been reduced and made equal to that of air holes of other rings. The decrease in the diameter leads to an increase in the leakage of the optical mode, as a result the optical mode does not exist in the core region and pushes out to the outer cladding region of the fiber. When the diameter $d_l$ is made larger in comparison to other circular air holes, the core cladding index contrast is sufficient to arrest the optical mode in the core region which has been demonstrated in figure 1(b). The birefringence of the fiber is contributed mainly due to the presence of elliptical air holes in the innermost ring. The birefringence of the fiber can be tailored by varying $\eta$, $d$ and air hole pitch. We have used silica glass with refractive index 1.45 as core. The mode field pattern of this fiber has been demonstrated in Fig. 1 (c). The mode field, which is elliptic, is confined in the core region.

### 2.3. Optical Properties of the Fiber

To begin with, we have evaluated birefringence of the fiber. The variation of birefringence with wavelength ($\lambda$) at different hole pitch has been displayed in Fig. 2. Fiber birefringence increases with the decrease in the value of fiber hole pitch $\Lambda$. At the telecommunication wavelength 1.55μm, the modal birefringence is very large (~$3.33 \times 10^{-2}$), which is the largest value of birefringence for index guiding PCF reported thus so far. The value of birefringence of the fiber can be tailored by varying the length of elliptical air holes and distance between air holes. Note that in a recent publication Mohit *et. al.* [33] have reported birefringence of $2.2 \times 10^{-2}$. The design of the present fiber closely resembles that of the reference [33]. However, in the present investigation, we have improved the design to achieve slightly higher value of birefringence. The wavelength dependencies of dispersion for horizontal and vertical polarization modes have been depicted in Fig. 3. The fiber exhibits normal dispersion above 1.3 $\mu m$ and anomalous dispersion below 1.3 $\mu m$. In the proposed fiber, two elliptical air holes with increased air filing fraction helps to increase the refractive index contrast between core and cladding, which influences the dispersion profile and other optical properties. From the Fig.3, we can see that the dispersion curve of vertical polarization mode is influenced more in comparison to the horizontal polarization mode. The core and cladding refractive index contrast in vertical direction causes the vibration in the dispersion profile between $0.5\ \mu m\ to\ 1\ \mu m$. Fig. 4. demonstrates the variation of the effective V-parameter of horizontal and vertical polarization modes for a typical hole pitch $\Lambda = 0.7 \mu m$. Effective V parameter decreases as wavelength increases. From the figure it is evident that $V_{eff} \leq 4.1$, hence, the fiber is endlessly single mode over a wide range of wavelength Fig. 5 depicts the variation of the walk off parameter with wavelength. Due to large birefringence, the value of a walk-off is very small. For a fixed value of hole pitch, the walk-off parameter initially increases smoothly with the increase in wavelength, then starts decreasing with the increasing $\lambda$. With the decrease in the value of pitch, the zero crossing shifts towards higher wavelength.

To study optical nonlinearity of the fiber, we first examine the mode field. Several design parameters control the mode field, for example $d, d_a, d_b, d_l$ and $\Lambda$ contribute to the size of the mode field. For a fixed set of values of $d$, $d_a$, $d_b$ and $d_l$, the core cladding index contrast is influenced by the hole pitch. Smaller hole pitch enhances index contrast, thus felicitating tighter mode field confinement in the core. Fig. 6 shows the variation of the effective optical mode area with wavelength for different fiber pitch. Mode area increases with the increase in the value of the wavelength. Therefore, the smaller hole pitch reduces mode area, thereby enhancing fiber nonlinearity. In the Fig. 7, the designed fiber exhibits large nonlinearity, for example $\gamma \approx 79\ W^{-1}km^{-1}$ at $1.06\mu m$ for, which is quite large for silica fibers. Such large optical nonlinearity will be useful in generating SC. As we decrease the core region, due to the high contrast between core and cladding the refractive index, optical mode confines which cause the effective mode area decrements, results high effective nonlinearity. We can see that as we increase the core region or air hole pitch the effective mode area increases, but the effective nonlinearity decreases with an increase in the air hole pitch.

## 3. Supercontinuum Generation

### 3.1. Numerical method

We now proceed to investigate the SC generation in the designed fiber whose optical properties have been described in previous sections. The propagation of optical pulse in this fiber can be modeled using modified nonlinear Schrödinger equation [83]:

$$\frac{\partial}{\partial z}A(z,T) = -\frac{\alpha(\omega)}{2}A(z,T) + \sum_{n\geq 2}\beta_n \frac{i^{n+1}}{n!}\frac{\partial^n}{\partial t^n}A(z,T)$$

$$+ i\gamma\left(1 + \frac{i}{\omega_0}\frac{\partial}{\partial T}\right)\int_{-\infty}^{\infty} R(T')\,|A(z,T-T')|^2 dT' \,, \tag{1}$$

where $A(z,t)$ is the slowly varying envelope of the electric field of the optical pulse, $\alpha$ is the frequency dependent loss, $\beta_n$ is the nth order dispersion at the centre frequency $\omega_0$, $R(T)$ is the Raman response function which may be defined as $R(T) = (1-f_r)\delta(T) + f_r h_r(T)$, $f_r = 0.18$ for silica material and $h_r$ is calculated using $h_r(T) = \frac{\tau_1^2 + \tau_2^2}{\tau_1 \tau_2^2} exp\left(\frac{-\tau}{\tau_2}\right) sin\left(\frac{-\tau}{\tau_1}\right)$, with $\tau_1 = 12.2\,fs$ and $\tau_2 = 32\,fs$ for silica material. Further, while writing above equation, we have assumed that the input pulse will either possess horizontal or vertical polarization during their launch in the fiber, thus the issue of cross phase modulation of between pulses of two different polarizations does not arise. The nonlinear Schrödinger equation can be solved using the *split-step Fourier method*, which is one of the most popular methods due to its simplicity and high accuracy. We can solve this equation by rewriting the generalized nonlinear Schrödinger equation in the form:

$$\frac{\partial A}{\partial z} = (\widehat{D} + \widehat{N})A \qquad (2)$$

where $\widehat{D}$ is an operator that takes care of all linear terms and losses within a linear medium and $\widehat{N}$ is an operator for all nonlinear effects, acting on the propagating pulse. These operators can be further expressed as

$$\widehat{D} = -\frac{\alpha}{2} + \sum_{n\geq 2} \beta_n \frac{i^{n+1}}{n!} \frac{\partial^n}{\partial t^n} \qquad (3)$$

$$\widehat{N} = i\gamma \left(1 + \frac{i}{\omega_0} \frac{\partial}{\partial T}\right) \int_{-\infty}^{\infty} R(T') \, |A(z, T - T')|^2 dT' \qquad (4)$$

When optical pulses are propagating along the PCFs, the dispersion and nonlinear effects act simultaneously along the length of the fiber. However, the split-step Fourier method is based on the assumption that over a small distance h, the dispersive and nonlinear effects act independently. This makes the entire problem very simple while encountering a negligible error. In this method, the propagation from $z$ to $z + h$ is performed in two steps, the first step only considers the linear effects while the second step, the nonlinear effects. The linear part is solved using a Fast Fourier transform (FFT) algorithm which requires a careful consideration of the step size $\Delta z$ along the fiber and the time resolution $\Delta t$ used for the temporal window. We have employed 8192 step/cm along the fiber length and $2^{12}$ points in the time and frequency domains for the FFT algorithm.

## 3.2. Simulation of SC Generation

In order to investigate supercontinuum generation in the designed fiber, we have used the dispersion profile of figure (3). We have taken one meter PCF, whose parameters are $\Lambda = 0.7$ µm, $d_l = \Lambda$, $d_a = 6 \times \Lambda$, $d_b = 0.5 \times \Lambda$, $d = 0.7 \times \Lambda$.. For horizontal polarization mode the values of different higher order dispersion terms at 1064 nm are $\beta_2 = -7.1756 \times 10^{-4}\ ps^2/km,$ $\beta_3 = -2.8744 \times 10^{-5}\ ps^3/km,$ $\beta_4 = 7.5205 \times 10^{-8}\ ps^4/km,$ $\beta_5 = 7.0575 \times 10^{-11}\ ps^5/km,$ $\beta_6 = -6.0902 \times 10^{-13}\ ps^6/km,$ $\beta_7 = -1.6589 \times 10^{-16}\ ps^7/km,$ $\beta_8 = 1.295 \times 10^{-17}\ ps^8/km,$ $\beta_9 = -2.3674 \times 10^{-19}\ ps^9/km$, whereas the values of higher order dispersion terms for vertical polarization mode at 1064nm are $\beta_2 = -5.4 \times 10^{-3}\ ps^2/km,$ $\beta_3 = 1.368 \times 10^{-5}\ ps^3/km,$ $\beta_4 = -1.1395 \times 10^{-8}\ ps^4/km,$ $\beta_5 = 3.9866 \times 10^{-11}\ ps^5/km,$ $\beta_6 = -1.5183 \times 10^{-13}\ ps^6/km,$ $\beta_7 = 9.5944 \times 10^{-16}\ ps^7/km,$ $\beta_8 = -1.8895 \times 10^{-17}\ ps^8/km$ and $\beta_9 = -1.7152 \times 10^{-19}\ ps^9/km.$ For numerical simulation, we have used sech laser pulses with peak power 1 kW, FWHM pulse duration 50fs and central wavelength 1064nm. Since, we have used only one meter long fiber, the loss is not expected to influence supercontinuum generation significantly.

In this section, we first examine the supercontinuum spectra generated by the input pump pulses whose polarization is along the horizontal direction. We numerically simulated supercontinuum spectra at different fiber length. The spectral and temporal evolution of the horizontally polarized input pulse has been displayed in figure (8) for four different fiber lengths. To capture additional information about the spectral broadening dynamics, in figure (9) we also plot spectral broadening phenomenon using density plot. In the density plot, both the spectral and temporal intensity have been plotted using logarithmic density scale truncated at -40dB relative to the maximum value. Such plot is very useful in the generation and evolution of low amplitude temporal components [39]. Initially, i.e., For example, at 25 cm length of the fiber, the spectral broadening is dominated by the SPM. The spectrum possesses a typical oscillatory structure which is always associated with SPM induced spectral broadening. This oscillatory structure is created by spectral in interference of different identical spectral components which are present at different temporal locations of the pulse. Subsequently, several other important processes such as solitons formation, intra-pulse Raman scattering and solitons self-frequency as well as dispersive wave generation comes into play. The spectrum at the end of the fiber has been depicted at the top panel. The short wavelength side reaches up to 500nm while the infrared portion extends almost 1950nm. The spectral spread at the end of 100cm fiber is around 1450nm. Red shifted pulse is more intense in comparison to the blue shifted peaks and the broadening increases with the increase in the propagation length.

We now proceed to examine the SC generation of the fiber using identical pulses which are polarized along the vertical direction. Figure (10) and (11) demonstrates the SC. Notable differences from the earlier case can be easily noticed. The spectral broadening is much wider in the present case. At the end of the 100cm fiber, it extends from 650nm in the short wavelength side to approximately 3000nm in the far infrared side, a broadening of 2350nm. The spectra is less uniform in comparison to the earlier case. Due to larger dispersion, temporal broadening of the pulse is also larger in comparison to earlier case.

## 4. Conclusion

In this paper, we have designed a highly birefringent photonic crystal fiber with hexagonal lattice of air holes and investigated its optical properties. The birefringence, dispersion characteristics, walk-off, effective index, the fiber V-parameter and mode field of the fundamental mode have been numerically investigated using the finite difference time domain (FDTD) method. The designed fiber promises very large birefringence ($3.33 \times 10^{-2}$) at 1.55 µm. The fiber has negative dispersion at the wavelength 1.55 µm and possesses small walk-off near telecommunication wavelength. The generated supercontinuum for horizontal polarized light is more superior in comparison to that of vertically polarized light. For horizontal polarized light, the spectral spread at the end of 100cm fiber is around 1450nm. Red shifted pulse is more intense in comparison to the blue shifted peaks and the broadening increases with the increase in the propagation length. The spectral broadening for vertically polarized light is much wider. At the end of the 100cm fiber, it extends from 650nm in the short wavelength side to approximately 3000nm in the far infrared side, a broadening of 2350nm. The spectra is less uniform in comparison to the earlier case. Due to larger dispersion, temporal broadening of the pulse is also larger in comparison to earlier case.

## 5. Acknowledgments

This work is supported by the Department of Science and Technology (DST), Government of India, through the R&D grant SR/S2/LOP-17/2010 and University Grants Commission, Bahadur Shah Zafar Marg, New Delhi –100001 through a major R&D project. The authors would like to thank DST and UGC for the assistances. One of the authors, Mohit Sharma would like to thank DST, Government of India, for providing Fellowship and Mr. N. Borgohain would like to thank UGC for the same.

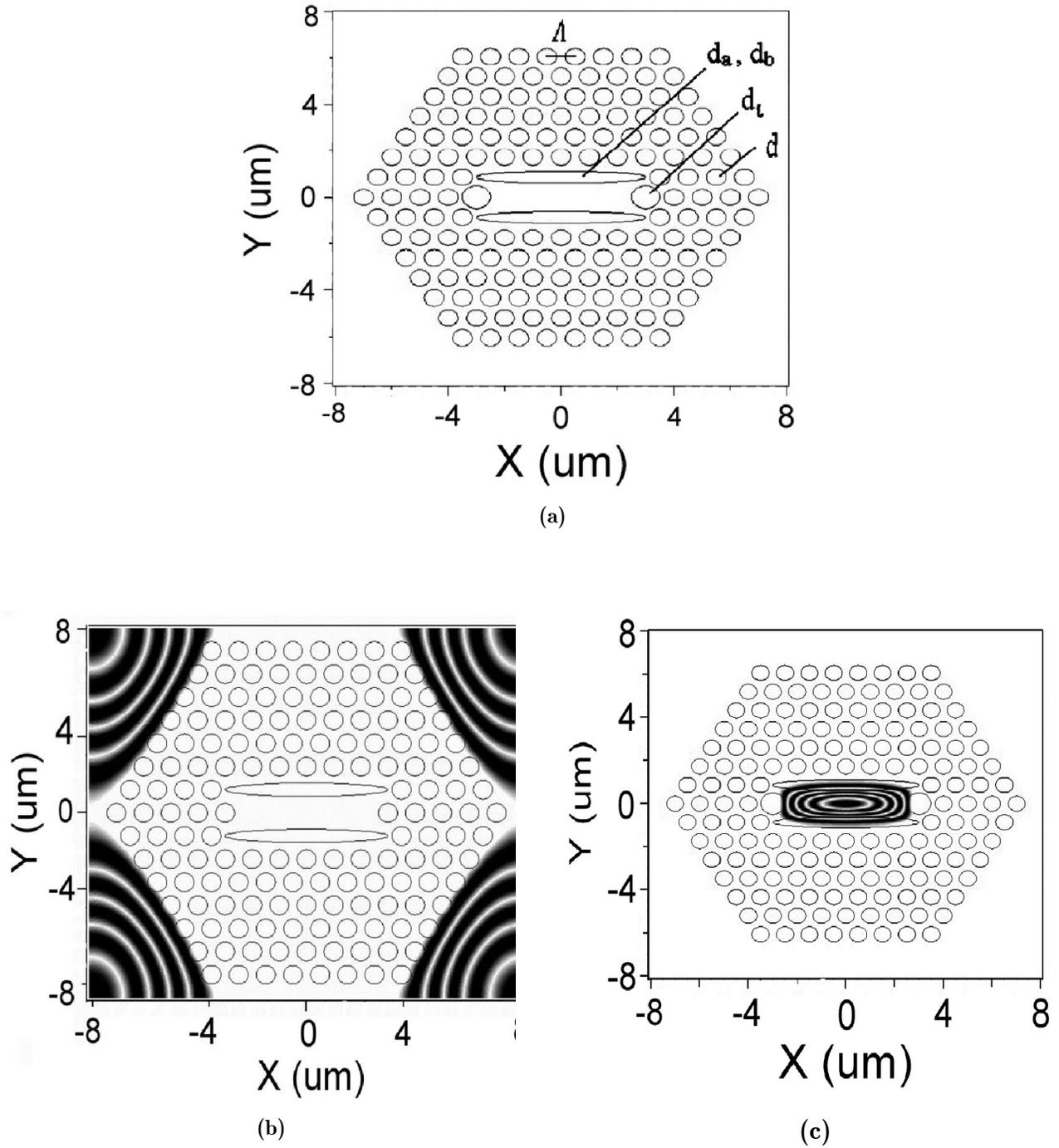

Fig. 1 (a) Schematic of the photonic crystal fiber with two big circular air holes of diameter $d_l$. (b) without big air holes, (c) Mode field of the fiber with big air holes. Hole pitch $\Lambda = 0.7$ μm, $d_l = \Lambda$, $d_a = 6 \times \Lambda$, $d_b = 0.5 \times \Lambda$, $d = 0.7 \times \Lambda$.

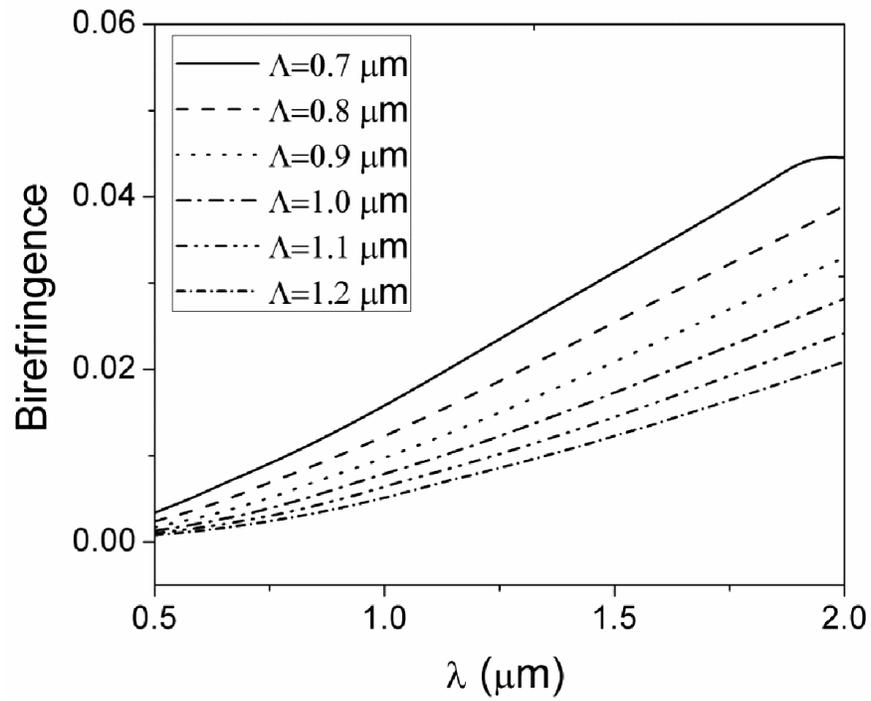

Fig. 2 Variation of fiber birefringence with wavelength for different hole pitch $\Lambda$, $d_l = \Lambda$, $d_a = 6 \times \Lambda$, $d_b = 0.5 \times \Lambda$, $d = 0.7 \times \Lambda$

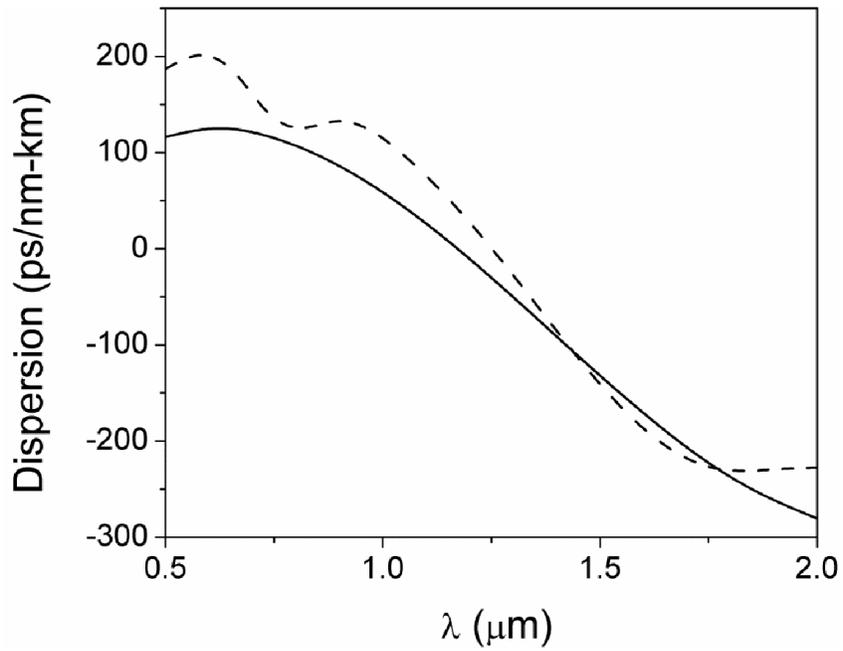

Fig. 3　Dispersion of horizontal and vertical polarization mode
　　　$\Lambda = 0.7\ \mu m,\ d_l = \Lambda,\ d_a = 6 \times \Lambda,\ d_b = 0.5 \times \Lambda,\ d = 0.7 \times \Lambda$.

Solid line represents horizontal polarization mode, while the dashed line signifies vertical polarization mode.

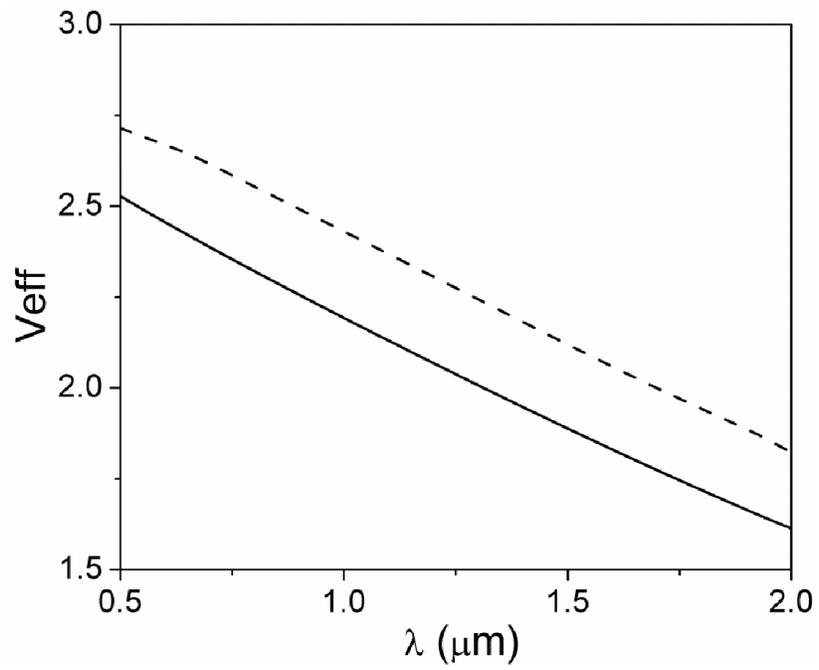

Fig. 4  Effective V Parameters of horizontal and vertical polarization mode

$\Lambda = 0.7$ μm, $d_l = \Lambda$, $d_a = 6 \times \Lambda$, $d_b = 0.5 \times \Lambda$, $d = 0.7 \times \Lambda$. Solid line represents horizontal polarization mode, while the dashed line signifies vertical polarization mode.

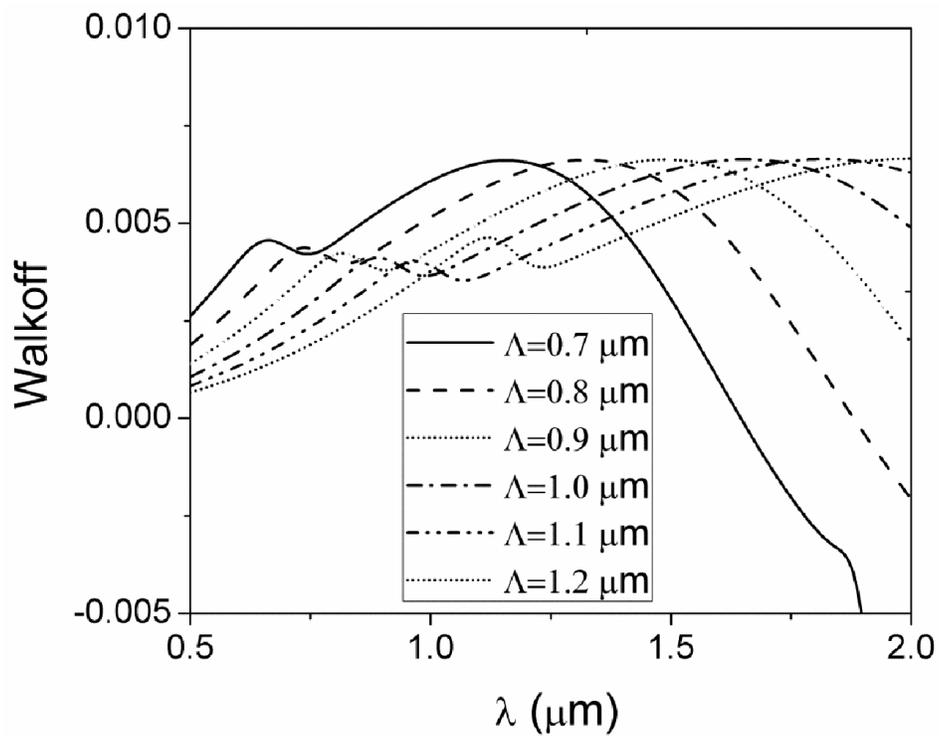

Fig.5 Variation of Walk-off with wavelength for different hole pitch $\Lambda$. $d_l = \Lambda$, $d_a = 6 \times \Lambda$, $d_b = 0.5 \times \Lambda$, $d = 0.7 \times \Lambda$.

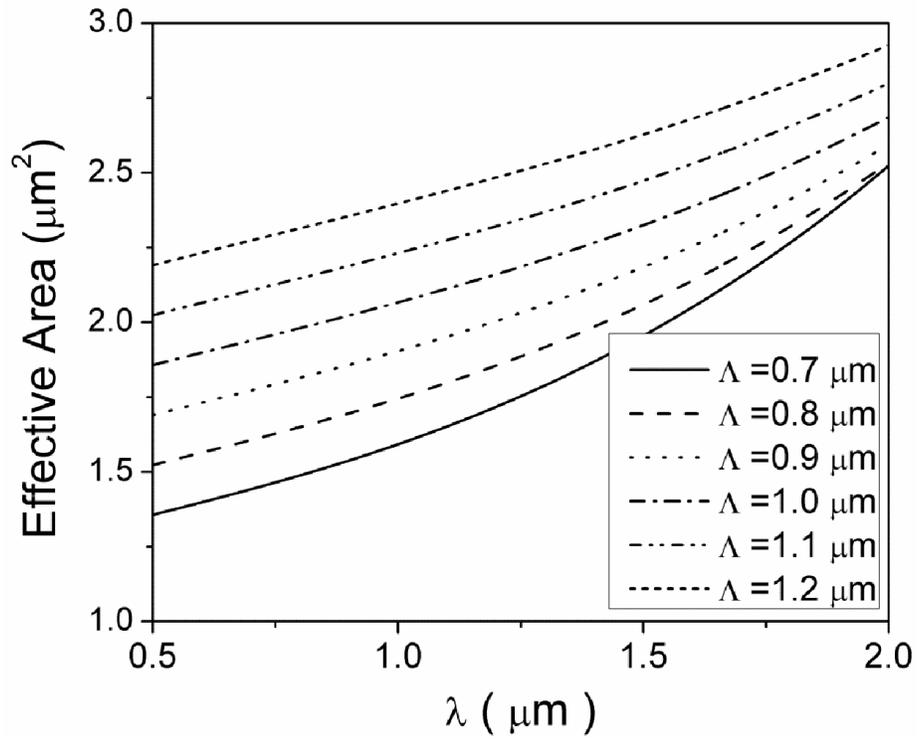

Fig. 6 Variation of Effective Area with wavelength for different hole pitch $\Lambda$. $d_l = \Lambda$, $d_a = 6 \times \Lambda$, $d_b = 0.5 \times \Lambda$, $d = 0.7 \times \Lambda$.

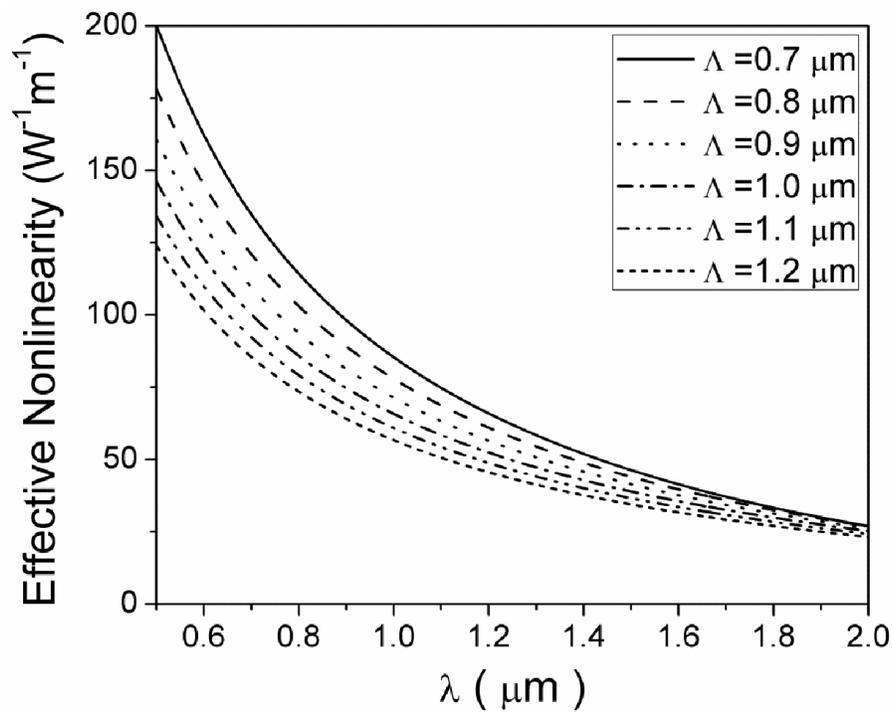

Fig. 7  Variation of Effective nonlinearity with wavelength for different hole pitch $\Lambda$. $d_l = \Lambda$, $d_a = 6 \times \Lambda$, $d_b = 0.5 \times \Lambda$, $d = 0.7 \times \Lambda$.

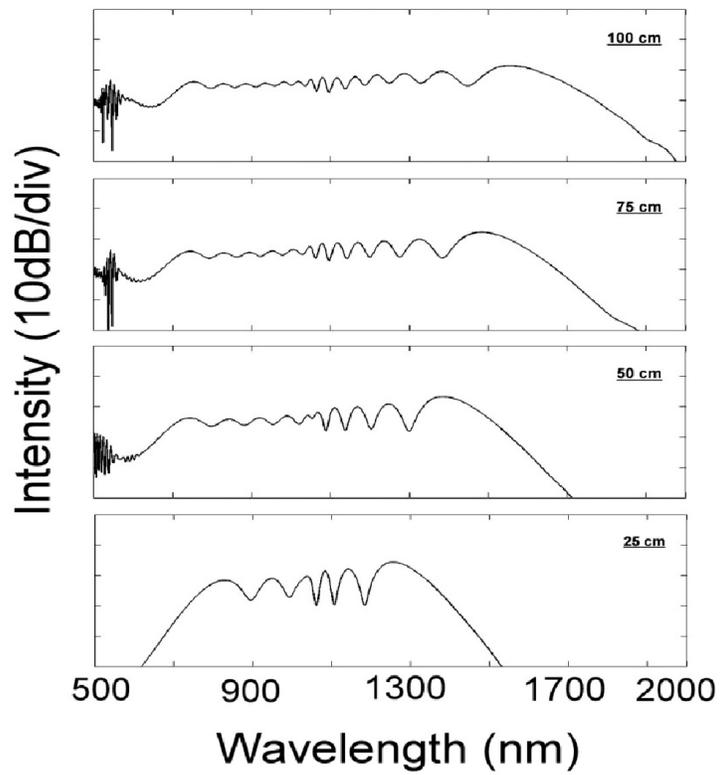

Fig.8(a) Spectra of supercontinuum generation by pumping a 1m long horizontal polarized silica PCF with $\Lambda = 0.7\,\mu m$, 50 fs pulse and 1 kW peak power.

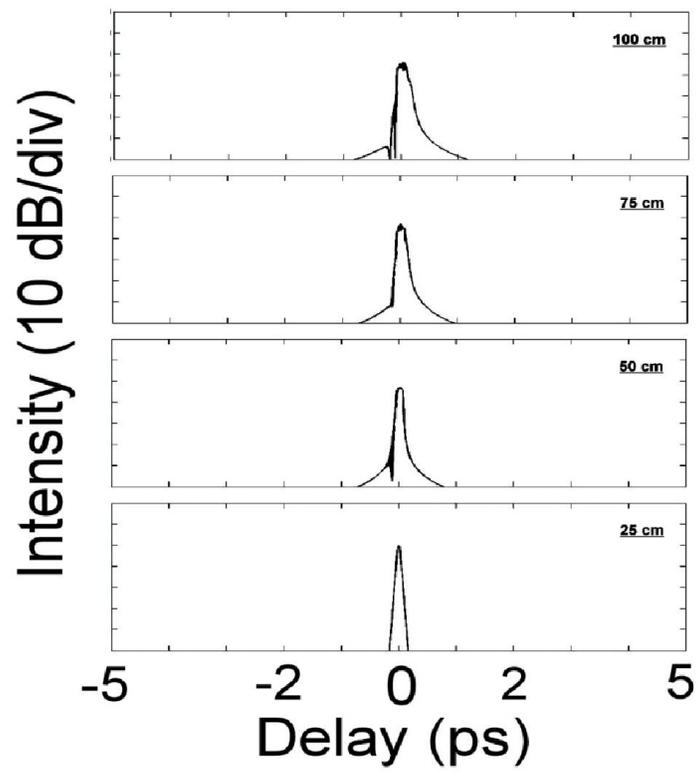

Fig.8(b) Temporal profile of supercontinuum generation by pumping a 1m long horizontal

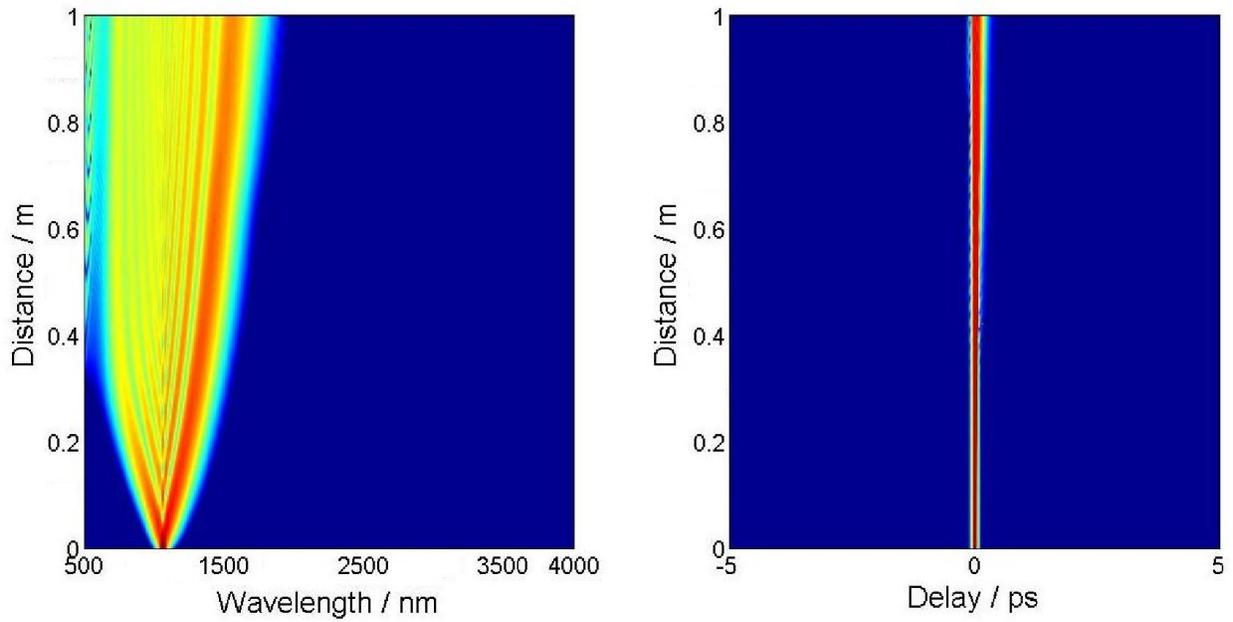

Fig. 9 Spectral and temporal evolution of supercontinuum generation by pumping a 1m long horizontal polarized silica PCF with Λ = 0.7 μm, 50 fs pulse and 1 kW peak power.

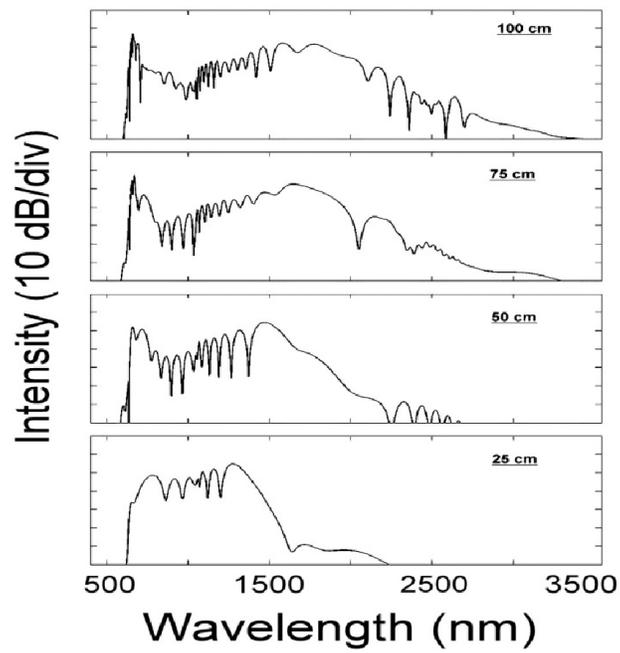

Fig.10(a) Spectra of supercontinuum generation by pumping a 1m vertical polarized long silica PCF with $\Lambda = 0.7\,\mu m$, 50 fs pulse and 1 kW peak power.

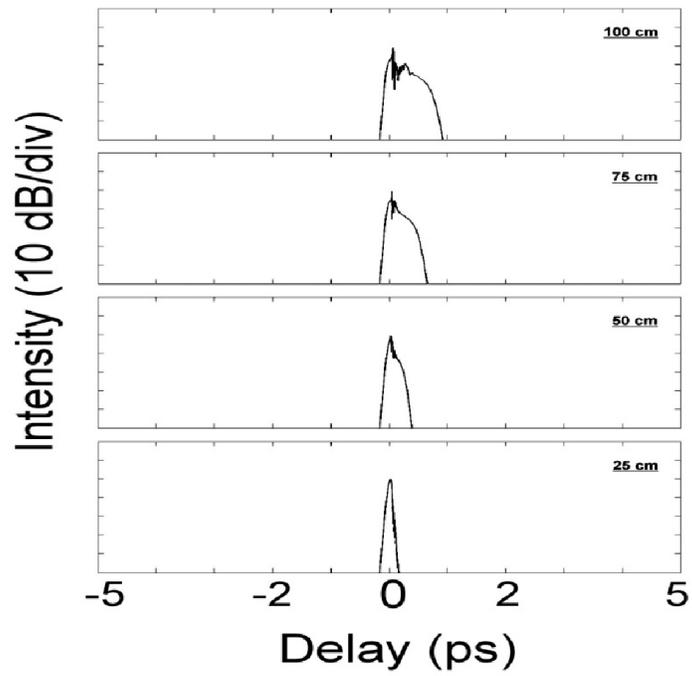

Fig.10(b) Temporal profile of supercontinuum generation by pumping a 1m long vertical polarized silica PCF with $\Lambda = 0.7$ μm, 50 fs pulse and 1 kW peak power.

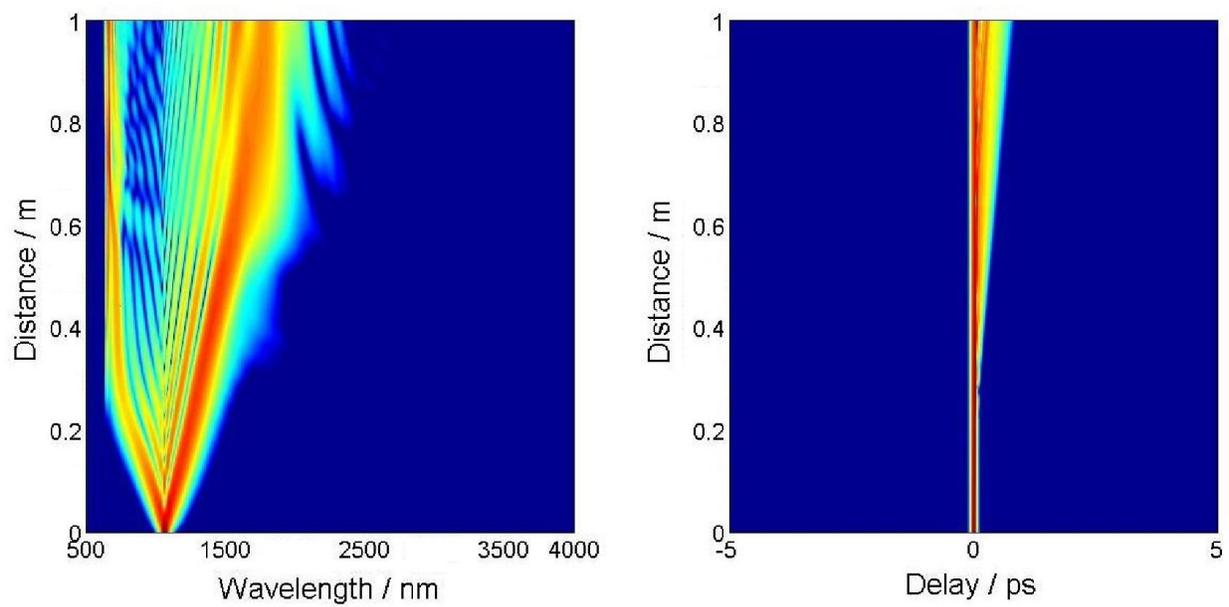

Fig. 11 Spectral and temporal evolution of supercontinuum generation by pumping a 1m long vertical polarized silica PCF with $\Lambda = 0.7\ \mu m$, 50 fs pulse and 1 kW peak power.